\documentclass[reprint,showpacs,preprintnumbers,amsmath,amssymb,longbibliography,aip]{revtex4-1}
\usepackage[utf8]{inputenc}
\usepackage{mathrsfs}
\usepackage{amsmath}
\usepackage{bm}
\usepackage{amsfonts}
\usepackage{amssymb}
\usepackage{amsthm}
\usepackage{color}
\usepackage{graphicx}
\graphicspath{{./images/}}
\usepackage{geometry}
\usepackage{hyperref}
\usepackage{ulem}
\usepackage{epstopdf}
\usepackage{graphicx}
\hypersetup{
     unicode=false,
     pdftoolbar=true,
     pdfmenubar=true,
     pdffitwindow=false,     % window fit to page when opened
     pdfstartview={FitH},    % fits the width of the page to the window
     pdftitle={My title},    % title
     pdfauthor={Author},     % author
     pdfsubject={Subject},   % subject of the document
     pdfcreator={Creator},   % creator of the document
     pdfproducer={Producer}, % producer of the document
     pdfkeywords={keyword1} {key2} {key3}, % list of keywords
     pdfnewwindow=true,      % links in new PDF window
     colorlinks=true,       % false: boxed links; true: colored links
     linkcolor=blue,          % color of internal links (change box color with linkbordercolor)
     citecolor=blue,        % color of links to bibliography
     filecolor=magenta,      % color of file links
     urlcolor=blue           % color of external links
}
\usepackage[toc,page]{appendix}

\begin{document}
\title{Spin-wave focusing induced skyrmion generation}
\author{Zhenyu Wang}
\affiliation{School of Electronic Science and Engineering and State Key Laboratory of Electronic Thin Films and Integrated Devices, University of Electronic Science and Technology of China, Chengdu 610054, China}
\author{Z.-X. Li}
\affiliation{School of Electronic Science and Engineering and State Key Laboratory of Electronic Thin Films and Integrated Devices, University of Electronic Science and Technology of China, Chengdu 610054, China}
\author{Ruifang Wang}
\affiliation{Department of Physics and Institute of Theoretical Physics and Astrophysics, Xiamen University, Xiamen 361005, China}
\author{Bo Liu}
%\email[ ]{liubo\_sm@163.com}
\affiliation{Key Laboratory of Spintronics Materials, Devices and Systems of Zhejiang Province, Hangzhou 311305, China}
\author{Hao Meng}
\affiliation{Key Laboratory of Spintronics Materials, Devices and Systems of Zhejiang Province, Hangzhou 311305, China}
\author{Yunshan Cao}
\affiliation{School of Electronic Science and Engineering and State Key Laboratory of Electronic Thin Films and Integrated Devices, University of Electronic Science and Technology of China, Chengdu 610054, China}
\author{Peng Yan}
\email[Corresponding author: ]{yan@uestc.edu.cn}
\affiliation{School of Electronic Science and Engineering and State Key Laboratory of Electronic Thin Films and Integrated Devices, University of Electronic Science and Technology of China, Chengdu 610054, China}

\date{\today}% It is always \today, today, but any date may be explicitly specified

\begin{abstract}
We propose a new method to generate magnetic skyrmions through spin-wave focusing in chiral ferromagnets. A lens is constructed to focus spin waves by a curved interface between two ferromagnetic thin films with different perpendicular magnetic anisotropies. Based on the principle of identical magnonic path length, we derive the lens contour that can be either elliptical or hyperbolical depending on the magnon refractive index. Micromagnetic simulations are performed to verify the theoretical design. It is found that under proper condition magnetic skyrmions emerge near the focus point of the lens where the spin-wave intensity has been significantly enhanced. A close investigation shows that a magnetic droplet first forms and then converts to the skyrmion accompanying with a change of topological charge.
Phase diagram about the amplitude and duration-time of the exciting field for skyrmion generation is obtained. Our findings would be helpful for designing novel spintronic devices combining the advantages of skyrmionics and magnonics.
\end{abstract}

\maketitle
Skyrmionics \cite{Nagaosa2013,Fert2013,Zhang201501,Krause2016,Zhang2020} and magnonics \cite{Schneider2008,Serga2010,Lenk2011,Chumak2015} are two emerging research fields in spintronics, which utilize skyrmions and spin waves (magnons when quantized) as carriers to encode, transmit, and process information, respectively. Magnetic skyrmions normally exist in chiral bulk magnets or magnetic thin films with the Dzyaloshinskii-Moriya interaction (DMI) \cite{Dzyaloshinsky1958,Moriya1960}. They are topologically protected spin textures and can not be nucleated and annihilated under continuous magnetization deformation. In contrast, magnons are the low-energy excitations in ordered magnets and can be created and destroyed due to their bosonic nature.

Realizing both skyrmion and magnon functionalities in a single spintronic device could significantly promote the development of magnetic memory and logic elements as an alternative to the conventional CMOS (complementary metal oxide semiconductor) computing technology. Indeed, the interaction between magnons and skyrmions has been extensively studied recently, such as magnon-skyrmion scattering \cite{Iwasaki2014,Schutte2014}, magnon-driven skyrmion motion \cite{Zhang2017,Jiang2020}, and skyrmion-based magnonic crystal \cite{Ma2015,Moon2016}. However, an important issue about the conversion between skyrmions and magnons has not been well investigated, although the spin-wave emission has been observed in the annihilation process or the core switching of magnetic skyrmion \cite{Zhang201702,Zhang2015}. A common view is that it is rather difficult to convert spin waves to magnetic skyrmions because the spin-wave energy is much lower than the barrier between the uniform ferromagnetic state and the skyrmion.

Over the past years, geometrical curvature effects in magnetism \cite{Streubel2016} have attracted considerable attention due to their promising application potential and rich physics, including the enhanced stability of domain wall \cite{Yan2010} and skyrmion \cite{Wang2019} in magnetic nanotubes, chirality symmetry breaking in ferromagnetic M\"{o}bius rings \cite{Pylypovskyi2015}, and magnonic Cherenkov-like effect \cite{Yan2011}, to name a few. Very recently, it has been demonstrated that the curved interface in two-dimensional ferromagnetic films can be used to construct a lens for spin-wave focusing \cite{Toedt2016,Bao2020}. The concept of spin-wave lens inspires us to accumulate energy to overcome the barrier for skyrmion formation.

\begin{figure}
  \centering
  \includegraphics[width=0.48\textwidth]{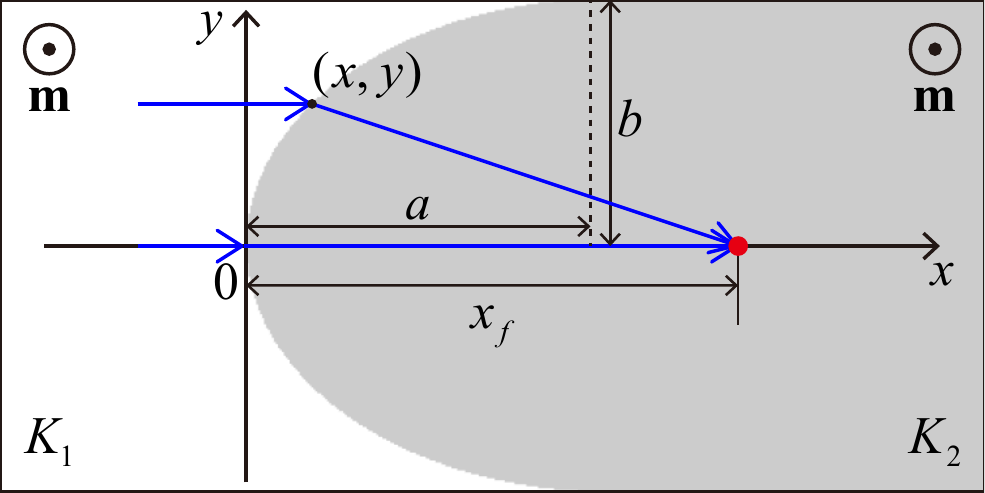}\\
  \caption{Schematic of a spin-wave lens with an elliptical interface between two ferromagnetic films with different perpendicular magnetic anisotropies $K_{1,2}$. The static magnetization $\mathbf{m}$ is oriented along $+z$ direction. The semi-major and semi-minor axes of the elliptical interface are $a$ and $b$, respectively. A parallel incident spin wave (blue arrows) propagates through the interface and converges on the focal point (red point).}\label{fig1}
\end{figure}

The idea is as follows: First, the spin-wave intensity near the focal point can be significantly enhanced, so that the nonlinear effect becomes dominating; Second, the focal-point magnetization oscillates strongly and might even be locally reversed, which is necessary for the nucleation of magnetic skyrmion. In this letter, we demonstrate that skyrmions can be created by spin-wave focusing in a magnetic film without introducing external defects as the nucleation sites \cite{Iwasaki2013,Mochizuki2017,Miyake2020}. To achieve a perfect focusing without spherical aberration \cite{Bao2020}, we design a spin-wave lens with a curved interface [see Eq. (\ref{eq_lens_contour}) below] between two ferromagnetic films with different perpendicular magnetic anisotropies [$K_{i}, (i=1,2)$], as shown in Fig. \ref{fig1}. The magnetic anisotropy change in this heterogeneous thin film can be realized by the recently discovered effect of the voltage-controlled magnetic anisotropy \cite{Amiri2012}. Initially, the heterogenous films are uniformly magnetized along $+\hat{z}$ direction. The spin-wave dynamics is described by the Landau-Lifshitz-Gilbert (LLG) equation,
\begin{equation}\label{eq_llg}
\frac{\partial\mathbf{m}}{\partial t}=-\gamma\mu_{0}\mathbf{m}\times\mathbf{H}_{\mathrm{eff}}+\alpha\mathbf{m}\times\frac{\partial\mathbf{m}}{\partial t},
\end{equation}
where $\mathbf{m}=\mathbf{M}/M_{s}$ is the unit magnetization vector with the saturated magnetization $M_{s}$, $\gamma$ is the gyromagnetic ratio, $\mu_{0}$ is the vacuum permeability, and $\alpha$ is the Gilbert damping constant.
The effective field $\mathbf{H}_{\mathrm{eff}}$ comprises the exchange field, the DM field, the anisotropy field, and the dipolar field.
The dipolar interaction is approximated by the demagnetization field $\mathbf{H}_{\mathrm{d}}=-M_{s}m_{z}\hat{z}$.
Neglecting the damping term ($\alpha=0$), the spin-wave spectrum can be obtained by solving the linearized LLG equation:
\begin{equation}\label{eq_dispersion}
  \omega(\mathbf{k})=A^{\ast}\mathbf{k}^{2}+\omega_{K_{i}},
\end{equation}
where $A^{\ast}=2\gamma A/M_{s}$ with the exchange constant $A$, $\omega_{K_{i}}=2\gamma K_{\mathrm{eff},i}/M_{s}$ with the effective anisotropy constant $K_{\mathrm{eff},i}=K_{i}-\mu_{0}M_{s}^{2}/2$, and $\mathbf{k}=(k_{x},k_{y})$ is the wave vector of spin wave.
From the dispersion relation (\ref{eq_dispersion}), one can see that the DMI has no effect when the magnetization is perpendicular to the film plane \cite{Cortes2013,Wang2018}.

We approximate the refractive index of spin waves as follows \cite{Jeong2011}:
\begin{equation}\label{eq_nsw}
  n_{\mathrm{sw}}=\frac{ck}{\omega(\mathbf{k})},
\end{equation}
where $c$ is the light speed in vacuum and $k=|\mathbf{k}|$ is the wave number of spin wave.
Thus, the relative refractive index of spin waves is defined as the ratio of the magnonic wave number in the right domain (gray region in Fig. \ref{fig1}) to that in the left one (white region in Fig. \ref{fig1}):
\begin{equation}\label{eq_n}
  n=\sqrt{\frac{\omega-\omega_{K_{2}}}{\omega-\omega_{K_{1}}}}.
\end{equation}
In analogy to optics, we define the magnonic path length (MPL) as the distance spin waves propagate multiplied by the refractive index.
A perfect lens can be designed based on the identical MPL principle.
We assume a parallel spin wave incident from the left and converges into a focal point $(x_{f},0)$ in the right, as shown in Fig. \ref{fig1}.
Then the MPL principle yields
\begin{equation}\label{eq_mpl}
  x+n\sqrt{(x_{f}-x)^{2}+y^{2}}=nx_{f},
\end{equation}
and the lens contour is described by
\begin{equation}\label{eq_lens_contour}
 \frac{(x-a)^{2}}{a^{2}}+\frac{y^{2}}{b^{2}}=1,
\end{equation}
where $a=\frac{n}{n+1}x_{f}$ and $b=\frac{\sqrt{n^{2}-1}}{n+1}x_{f}$. One can see that the lens shape is elliptical for $n>1$ and hyperbolical for $n<1$. Here, we focus on the $n>1$ case, as depicted in Fig. \ref{fig1}.

\begin{figure}
  \centering
  \includegraphics[width=0.48\textwidth]{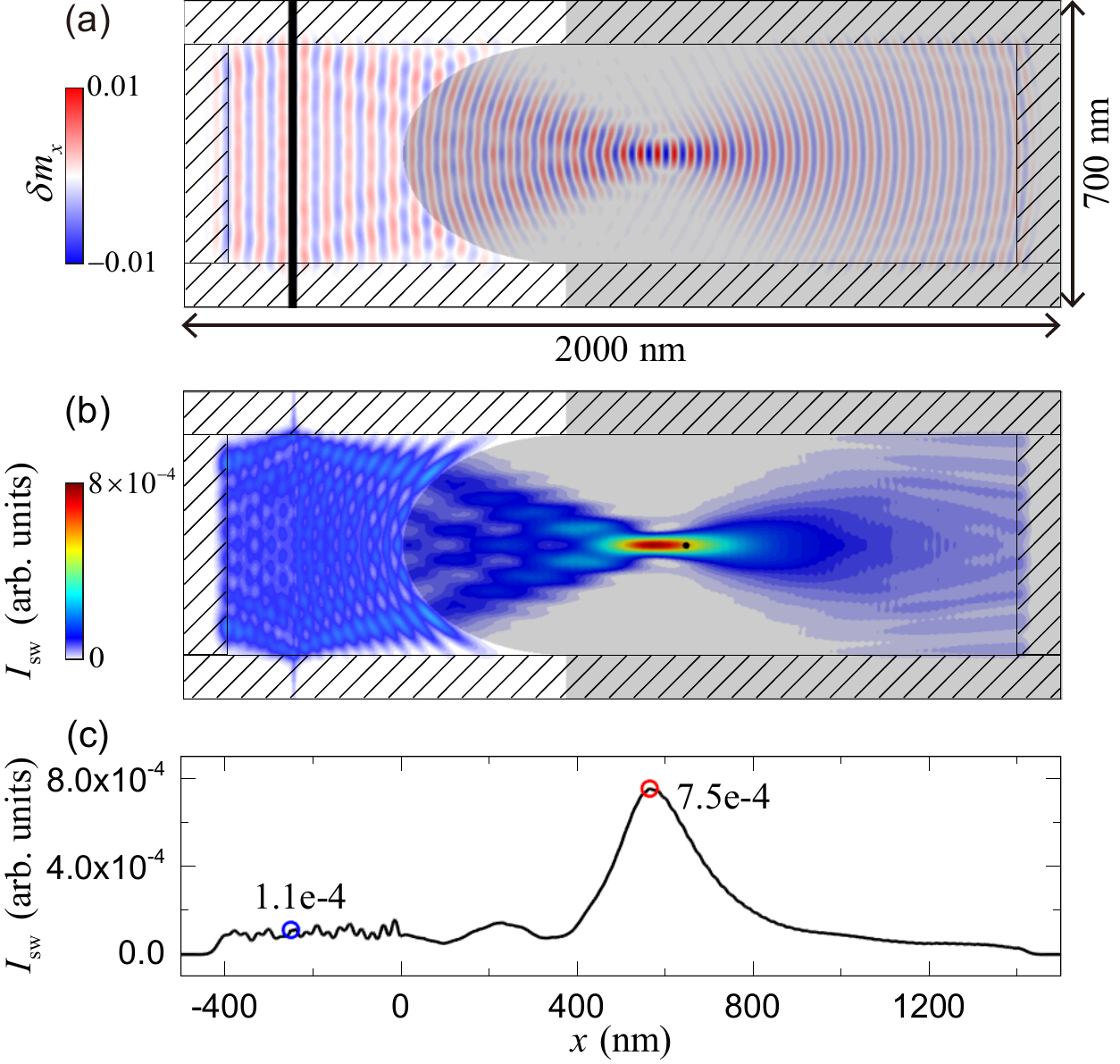}\\
  \caption{(a) Snapshot of spin wave with $\mu_{0}h_{0}=10$ mT and $\omega/2\pi=80$ GHz across the elliptical interface. The black bar denotes the exciting source of spin waves. (b) Intensity of spin waves in (a). The black point represents the theoretical position of the focal point. Absorbing boundary conditions are adopted in the dashed region in (a) and (b). (c) The profile of the spin-wave intensity along $x$ axis at $y=0$ nm in (b). The blue and red open dots indicate the spin-wave intensity at the exciting source and the focal point, respectively.}\label{fig2}
\end{figure}

To verify our theoretical design, micromagnetic simulations are performed using MuMax3 \cite{Vansteenkiste2014}.
We consider a heterogeneous magnetic thin film with the length 2000 nm, width 700 nm, and thickness 1 nm. The cell size of $2\times2\times1$ $\mathrm{nm^{3}}$ is used in simulations.
Magnetic parameters of Co are adopted in simulations \cite{Sampaio2013}: $M_{s}=5.8\times10^{5}$ $\mathrm{A/m}$, $A_{ex}=15$ $\mathrm{pJ/m}$, $D=2.5$ $\mathrm{mJ/m^{2}}$, $K_{1}=8\times10^{5}$ $\mathrm{J/m^{3}}$, and $K_{2}=6\times10^{5}$ $\mathrm{J/m^{3}}$.
In the dynamic simulations, a Gilbert damping constant of $\alpha=0.001$ is used to ensure a long-distance propagation of spin waves, and absorbing boundary conditions are adopted in the dashed area in Figs. \ref{fig2}(a) and \ref{fig2}(b) to avoid the spin-wave reflection by the film edges \cite{Venkat2018}.

We apply a sinusoidal monochromatic microwave field $\mathbf{H}_{\mathrm{ext}}=h_{0}\sin(\omega t)\hat{x}$ in a narrow rectangular area [black bar in Fig. \ref{fig2}(a)] to excite the incident spin waves.
We set the amplitude and frequency of the oscillating field as $\mu_{0}h_{0}=10$ mT and $\omega/2\pi=80$ GHz.
Based on Eq. (\ref{eq_n}), it is found that the relative refractive index is frequency-dependent and is $n=1.35$ for 80 GHz.
We set the semi-minor axis of the elliptical interface as $b=250$ nm and the corresponding $a$ and $x_{f}$ can be calculated as 370 and 644 nm, respectively.
Numerical results from magnetic simulations are shown in Fig. \ref{fig2}(a).
We also calculate the spin-wave intensity using the equation $I_{\mathrm{sw}}(x,y)=\int_{0}^{t}[\delta m_{x}(x,y,t)]^{2}dt$, as plotted in Fig. \ref{fig2}(b).
In Figs. \ref{fig2}(a) and \ref{fig2}(b), we can observe a significant focusing effect of spin waves in the right domain.
However, it is noted that the focus point obtained from the numerical simulation is shifted along $-x$ from the ideal position [black point shown in Fig. \ref{fig2}(b)].
It is due to the ray optic approximation for analyzing the spin-wave propagation, which requires the wavelength of spin waves much smaller than the lens size ($\lambda\ll a,b$).
One can diminish such deviation by increasing the size of the interface lens with respect to $\lambda$.
In Fig. \ref{fig2}(c), the profile of the spin-wave intensity along $x$ axis at $y=0$ nm is plotted. It shows that the spin-wave intensity has been enhanced by one order of magnitude, which is comparable with that of the graded index lens \cite{Whitehead2018,Vogel2020}.
It is worthy mentioning that the focus position $x_{f}$ can be tuned by varying the excitation frequency or by applying a perpendicular static magnetic field \cite{Bao2020}. In such cases, additional spherical aberration may emerge, unless the lens shape is reconstructed based on (\ref{eq_lens_contour}).

\begin{figure}
  \centering
  \includegraphics[width=0.48\textwidth]{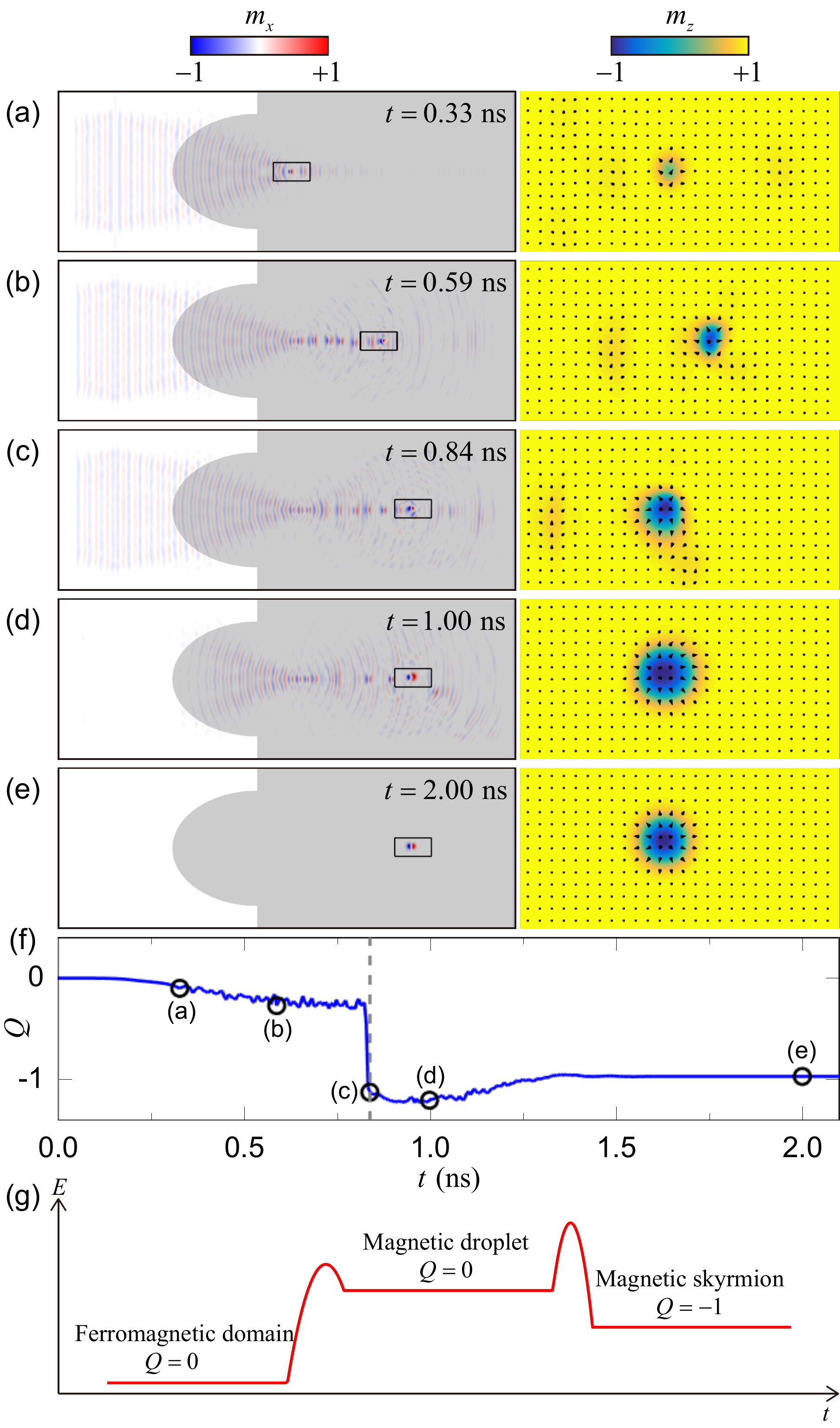}\\
  \caption{The creation process of magnetic skyrmion induced by the spin-wave focusing with the DMI ($D=2.5$ $\mathrm{mJ/m^{2}}$). The exciting field with $\mu_{0}h_{0}=350$ mT is applied in (a)-(c) and is turned off in (d)-(e).
  The left column shows the x-component magnetization of the heterogeneous film. The z-component magnetization of the rectangular areas in the left column is enlarged in the right column. (f) Temporal evolution of the topological number $Q$. The microwave field starts at $t=0$ and ends at $t=0.84$ ns indicated by the gray dashed line. (g) Schematic diagram of energy analysis of ferromagnetic domain, magnetic droplet, and magnetic skyrmion.}\label{fig3}
\end{figure}

To create skyrmion, we increase the exciting field amplitude to $\mu_{0}h_{0}=350$ mT.
A strong magnetization oscillation is observed around the focal point, as shown in Fig. \ref{fig3}(a).
With the continuous excitation of spin waves, more energies are harvested leading to the formation of magnetic droplet, which can be easily driven by spin waves [see Fig. \ref{fig3}(b)].
Magnetic droplet is a strongly nonlinear and localized spin-wave soliton \cite{Hoefer2010,Mohseni2013}.
In a chiral magnetic film, the trivial magnetic droplet is unstable due to the high DMI energy.
Assisted by spin waves, magnetic droplet is converted to a dynamical skyrmion at $t=0.84$ ns, as shown in Fig. \ref{fig3}(c).
We then turn off the microwave field and the system is relaxed toward an equilibrium state [see Fig. \ref{fig3}(d)].
After 1 ns, a stable skyrmion is formed, as shown in Fig. \ref{fig3}(e).
Figure \ref{fig3}(f) plots the time evolution of the topological charge $Q=(1/4\pi)\iint\mathbf{m}\cdot(\partial_{x}\mathbf{m}\times\partial_{y}\mathbf{m})dxdy$.
We observe an abrupt change of $Q$ at 0.84 ns, which provides a further evidence of the skyrmion creation.

We emphasize that, in creating skyrmions (see supplementary material Video 1), magnetic droplet acts an indispensable intermediate between ferromagnetic and skyrmion states. Although skyrmion is more stable than droplet, the energy barrier between ferromagnetic state and skyrmion is higher than that between ferromagnetic state and droplet, as shown in Fig. \ref{fig3}(g).
This is because the ferromagnetic state can be transformed to magnetic droplet in a continuous fashion and without changing the topological number ($Q=0$).
However, the change of $Q$ is needed for the transition from homogeneous ferromagnetic state to textured skyrmion, which requires more energy input from the external field.

\begin{figure}
  \centering
  \includegraphics[width=0.48\textwidth]{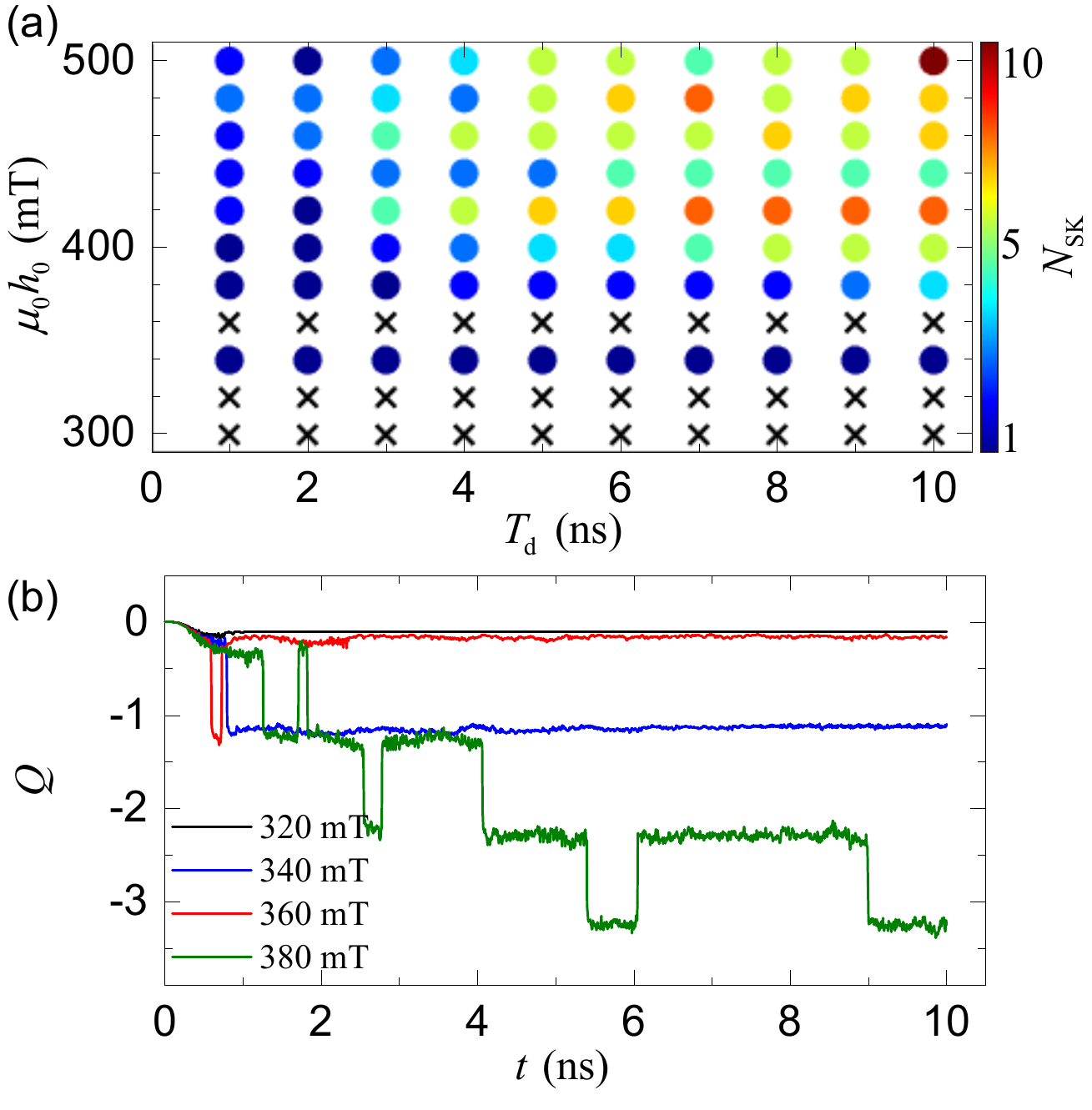}\\
  \caption{(a) Phase diagram of skyrmion creation with respect to the amplitude $h_{0}$ and duration time $T_{d}$ of the exciting field. The black crosses denote no skyrmion creation and the color dots represent the number of created skyrmions. (b) Temporal evolutions of the topological charge $Q$ under different field amplitudes.}\label{fig4}
\end{figure}

In Fig. \ref{fig4}(a), phase diagram of skyrmion creation induced by spin-wave focusing is shown.
It was expected that the number $N_{\text{sk}}$ of the generated skyrmions should increases with the amplitude $h_{0}$ and duration time $T_{d}$ of the microwave field.
However, the simulation results do not strictly follow this expectation.
For example, spin waves excited by the microwave field with $\mu_{0}h_{0}=340$ mT can produce one skyrmion, while no skyrmion is created under the exciting field with $\mu_{0}h_{0}=360$ mT.
To figure out the reason, we plot the time evolution of the topological charge $Q$ in Fig. \ref{fig4}(b).
The skyrmion generation and annihilation can be confirmed by the increase and decrease of $|Q|$, respectively.
One can see that the topological charge $Q$ is always around 0 for 320 mT and this indicates no skyrmion creation in the whole processes.
For 340 mT, one abrupt change of $Q$ from 0 to -1 at $t=0.85$ ns is observed, which represents the creation of a skyrmion.
For 360 mT, the topological charge $Q$ is around $-1$ only in a short period from 0.6 to 0.72 ns and vanishes then.
Under a higher exciting field amplitude 380 mT, we can observe multiple changes of $Q$, corresponding to more skyrmion creation and annihilation.
A close observation of the skyrmion evolution shows that the skyrmion annihilation is induced by the interaction between magnetic droplet and skyrmion: After the skyrmion formation, magnetic droplets are still continuously generated by the lens and they crash into a skyrmion, leading to a new droplet or a skyrmion annihilation accompanied by spin-wave emissions (see supplementary material Video 2).

The results reported here do not depend on the type of the spin-wave lens.
Other spin-wave lens, such as graded index lenses \cite{Dzyapko2016,Whitehead2018,Vogel2020} and magnonic meta-lenses \cite{Zelent2019,Grafe2020}, can also focus spin waves for the skyrmion generation.
Furthermore, our design can also be utilized to generate the Bloch-type skyrmion which is stabilized in the presence of the bulk DMI \cite{Bak1980}.
It should be mentioned that there have been several important proposals about the skyrmion creation using magnetic microwave field \cite{Mochizuki2017,Miyake2020,Flovik2017,Li2017}, electric current \cite{Iwasaki2013,Everschor2017,Buttner2017,Woo2018}, and others \cite{Koshibae2014,Zhou2014,Liu2015,Nii2015,Polyakov2020}.
However, creating skyrmions at a specific place using magnetic fields is not straightforward, because locally applying magnetic field over a nanoscopic region is challenging.
This problem is conventionally resolved by fabricating a nanoscopic defect on the ferromagentic thin film \cite{Mochizuki2017,Miyake2020}, which however poses new issues for devices performance and scalability.
For the skyrmion creation by electric current, the Joule heating due to the excessive high critical current density is a bottleneck. Our proposal in this work well avoids these problems.

In summary, we theoretically investigated the generation of magnetic skyrmion induced by spin-wave focusing. The lens contour was derived based on the identical magnonic path length principle. Micromagnetic simulations were performed to demonstrate the effective focusing of spin waves through the curved interface. By increasing the amplitude of the exciting field, strong nonlinear effects emerge close to the focus of the spin-wave lens. We observed spin-wave focusing induced magnetic skyrmion nucleation, mediated by unstable magnetic droplets. Our findings provide a new method to create skyrmions in magnetic thin film without artificial defects and would promote the development of spintronic devices combining spin waves and skyrmions.

%\section*{supplementary material}
See supplementary material for animations showing the skyrmion generation. The amplitude of excitation field $h_{0}=350$ and 400 mT in Video 1 and 2, respectively.
%\section*{acknowledgments}

We thank Z. Zhang for helpful discussions.
This work is supported by the National Natural Science Foundation of China (NSFC) (Grants No. 11604041 and 11704060) and the National Key Research Development Program under Contract No. 2016YFA0300801. Z.W. acknowledges the financial support from the China Postdoctoral Science Foundation under Grant No. 2019M653063. Z.-X.L. acknowledges the financial support of the China Postdoctoral Science Foundation (Grant No. 2019M663461) and the NSFC Grant No. 11904048.

\section*{DATA AVAILABILITY}
The data that support the findings of this study are available
from the corresponding author upon reasonable request.

\end{document}